\documentclass[amsmath, showpacs, twocolumn, superscriptaddress, aps]{revtex4}

\usepackage{txfonts}
\usepackage{bm}
\usepackage{graphicx,amsmath,mathrsfs}
\usepackage{amssymb}
\usepackage{epstopdf}
\usepackage{float}
\usepackage[section]{placeins}
\usepackage{tabularx}
\usepackage{multirow}
\usepackage{color}
\usepackage{ifpdf}
\usepackage{graphicx, latexsym, CJK, bbm}
\usepackage{amsfonts}
\usepackage{booktabs}
\usepackage{enumerate}
\usepackage{slashed}
\usepackage{hyperref}
\newcommand{\beq}{\begin{equation}}
\newcommand{\eeq}{\end{equation}}
\newcommand{\beqn}{\begin{eqnarray}}
\newcommand{\eeqn}{\end{eqnarray}}
\newcommand{\bea}{\begin{array}}
\newcommand{\eea}{\end{array}}
\newcommand{\bsub}{\begin{subequations}}
\newcommand{\esub}{\end{subequations}}
\newcommand{\bpm}{\begin{pmatrix}}
\newcommand{\epm}{\end{pmatrix}}

\newcommand{\scr}[1]{{\mathscr #1}}
\newcommand{\ff}[1]{\frac{1}{#1}}

\newcommand{\lrl}[1]{\left|#1\right|}
\newcommand{\lrc}[1]{\left<#1\right>}
\newcommand{\lrlc}[1]{\left|#1\right>}

\newcommand{\lrb}[1]{\Big(#1\Big)}
\newcommand{\lrs}[1]{\Big[#1\Big]}

\newcommand{\svec}[1]{{\mbox{\boldmath${ #1}$}}}
\newcommand{\ivec}{\vec}

\newcommand{\re}{\nonumber\\}

\newcommand{\sigs}{{\sigma\text{-S} }}
\newcommand{\omev}{{\omega\text{-V} }}
\newcommand{\rhov}{{\rho  \text{-V} }}
\newcommand{\rhot}{{\rho  \text{-T} }}
\newcommand{\rhvt}{{\rho  \text{-VT}}}
\newcommand{\pipv}{{\pi   \text{-PV}}}
\newcommand{\couv}{{ A    \text{-V} }}

\usepackage{ulem}
\newcommand{\delete}{\bgroup\markoverwith{\textcolor{red}{\rule[0.5ex]{2pt}{1pt}}}\ULon}

\renewcommand{\insert}[1]{\textcolor[rgb]{1.00,0.00,0.00}{#1}}

\allowdisplaybreaks

\begin{document}
\begin{CJK*}{GBK}{song}

\title{Self-consistent tensor effects on nuclear matter system under relativistic Hartree-Fock approach}

\author{Li Juan Jiang}
\affiliation{School of Nuclear Science and Technology, Lanzhou University, Lanzhou 730000, China}
\author{Shen Yang}
\affiliation{School of Nuclear Science and Technology, Lanzhou University, Lanzhou 730000, China}
\author{Jian Min Dong}
\affiliation{Institute of Modern Physics, Chinese Academy of Sciences, Lanzhou 730000, China}
\author{Wen Hui Long}
\email{longwh@lzu.edu.cn}
\affiliation{School of Nuclear Science and Technology, Lanzhou University, Lanzhou 730000, China}

\begin{abstract}
With the relativistic representation of the nuclear tensor force that is included automatically by the Fock diagrams, we explored the self-consistent tensor effects on the properties of nuclear matter system. The analysis were performed within the density-dependent relativistic Hartree-Fock (DDRHF) theory. The tensor force is found to notably influence the saturation mechanism, the equation of state and the symmetry energy of nuclear matter, as well as the neutron star properties. Without introducing any additional free parameters, the DDRHF approach paves a natural way to reveal the tensor effects on the nuclear matter system.

\end{abstract}


\maketitle

\end{CJK*}

\section{Introduction}
In the past several decades, the covariant density functional theories have achieved great successes in exploring the finite nuclei and nuclear matter. One of the most outstanding schemes is the relativistic mean field (RMF) theory with a limited number of free parameters \cite{Walecka:1974, Serot:1986, Reinhard1989, Ring1996, Bender:2003, Meng2006, Niksic2011Prog.Part.Nucl.Phys.519}. Because of its covariant formulation of strong scalar and vector fields, the RMF theory is able to self-consistently describe the nuclear spin-orbit effect. However, important degrees of freedom associated with the $\pi$ and tensor-$\rho$ fields are missing in the limit of Hartree approach. In fact, the dominant part of one-pion exchange process is the nuclear tensor force component \cite{Otsuka2005, Long2008} that plays significant roles in nuclear structure \cite{Colo2007, Long2008, Lalazissis2009PRC041301}, excitation and decay modes \cite{Bai2009PLB, Bai2009PRC, Bai2010PRL, Bai2011O16, Bai2011PRC}, and symmetry energy \cite{ Baoan2010, Vidana2011}.

As an important ingredient of nuclear force, the tensor force, together with the spin-orbit coupling, characterizes the spin dependent feature \cite{Otsuka2005}. It was firstly recognized by the discovery of electric quadrupole moment of the deuteron \cite{Fayache1997}. From the viewpoint of the meson exchange picture of the nucleon-nucleon interaction \cite{Yukawa:1935}, the nuclear tensor force was thought to be originated from the exchange processes of the $\pi$ and tensor-$\rho$ fields, corresponding to the long and short range parts, respectively \cite{Otsuka2005, Fayache1997}. In general, the nuclear tensor force is identified by the following form,
\begin{align}\label{Tensor_Wigner}
S_{12}=& 3\lrb{\svec\sigma_1\cdot\svec q} \lrb{\svec\sigma_2\cdot\svec q} - \svec\sigma_1\cdot\svec\sigma_2\svec q^2,
\end{align}
where $S_{12}$ is a rank-2 irreducible tensor well defined in the non-relativistic quantum mechanics, with the momentum transfer $\svec q=\svec p_1-\svec p_2$. In recent years, the nuclear tensor force was shown to play an essential role in determining the shell evolution from the stable to exotic nuclear systems, either by the nonrelativistic or relativistic calculations \cite{Otsuka2005, Colo2007, Lesinski2007, Long2007, Zuo2008, Otsuka2010, Wang2013}, although some suspicious remains due to the fact that the particle-vibration couplings were not included \cite{Afanasjev2014arxiv}. Furthermore, the inclusion of the nuclear tensor force also brought substantial impact on understanding the nature of the nuclear excitations and decay modes \cite{Cao2009PRC, Bai2009PLB, Bai2010PRL, Minato2013}. For the density-dependent behavior of nuclear symmetry energy --- the key quantity to understand the nuclear equation of state (EoS) and relevant astrophysical processes \cite{Baoan2008, Lattimer2004}, the tensor effects have also been revealed to be among the physics responsible for the uncertainty of the symmetry energy at supranuclear densities \cite{Baoan2010, Vidana2011}.

Although the nuclear tensor force has been well identified with the form \insert{(\ref{Tensor_Wigner})}, researchers encounter some difficulties due to the evident model dependence in determining its coupling strength based on the well-developed energy functionals such as the Skyrme forces \cite{Sagawa2014Colo}. Within the covariant density functional theory founded on the meson exchange picture of the nuclear force, people attempt to investigate the tensor effects by including the Lorentz tensor couplings, e.g., in terms of $\omega$-tensor couplings \cite{Mao2003}. However, such so-called "tensor" is just pure central-type contributions in the limit of Hartree approach. The solution is to introduce explicitly the Fock diagrams of the meson-nucleon couplings, so that the degrees of freedom associated with the $\pi$ and tensor-$\rho$ fields can be efficiently taken into account, for instance by the density-dependent relativistic Hartree-Fock (DDRHF) theory \cite{Long640, Long2010, Long2007}. Within DDRHF, substantial improvements due to the tensor effects have been revealed in reproducing the shell evolution without additional adjusted parameters \cite{Long2007, Long2008, Wang2013}. What is more, the relativistic representation of the nuclear tensor force was proposed very recently, with the new origin associated with the Fock diagrams of the isoscalar scalar $\sigma$- and vector $\omega$- couplings \cite{Jiang2014}. It has been confirmed that the spin-dependent feature --- the nature of the nuclear tensor force --- can be extracted and quantified almost completely by the proposed relativistic formalism \cite{Jiang2014}.

In this work, we will study the effects of the nuclear tensor force components which hide in the Fock diagrams of the meson-nucleon couplings, particularly the isoscalar scalar $\sigma$- and vector $\omega$-couplings, on the properties of nuclear matter and neutron stars. Section \ref{Sec:Theory} briefly introduces the relativistic formalism of the nuclear tensor forces for nuclear matter. In Sec. \ref{Sec:Result} are presented the calculated results and discussions, including the tensor effects on the bulk properties of symmetric nuclear matter and the EoS in Sec. \ref{Sec:Result-EOS}, on the density-dependent behavior of the symmetry energy in Sec. \ref{Sec:Result-Esym}, and on the neutron star structure in Sec. \ref{Sec:Result-NS}. Finally, a summary is given in Sec. \ref{Sec:Sum}.

\section{relativistic formalism of tensor force components in nuclear matter}\label{Sec:Theory}

\subsection{RHF energy functional}

Relativistically, the nucleon-nucleon ($NN$) interactions can be established on the picture of the meson exchanges, including the isoscalar and isovector ones. Consistent with this criterion, the Lagrangian density, i.e., the starting point of the relativistic Hartree-Fock (RHF) theory, can be constructed by enclosing the degrees of freedom of nucleon ($\psi$), two isoscalar mesons (scalar $\sigma$ and vector $\omega_\mu$), two isovector ones (pseudo-scalar $\ivec\pi$ and vector $\ivec\rho_\mu$), and photon ($A_\mu$) fields \cite{Bouyssy1987, Long639, Long2007}. Namely the $\sigma$- and $\omega$-meson fields are introduced to simulate the strong mid-range attraction and short-range repulsion, respectively, and the isovector part is evaluated by the $\pi$- and $\rho$-meson fields, and the photons take the Coulomb effects into account.

In general, the Lagrangian density $\scr L$ is composed of two parts, the free Lagrangian $\scr L_0$ and the one $\scr L_I$ describing the interactions between the nucleons and mesons (photons),
\begin{align}
\scr L=& \scr L_0 + \scr L_I,\label{Lagrangian}\\
\scr L_0=&\bar\psi\lrb{i\gamma_\mu\partial^\mu - M}\psi\label{Lfree}\\
&+ \ff2\partial_\mu\sigma\partial^\mu\sigma - \ff2 m_\sigma^2\sigma^2 +\ff2 m_\omega^2 \omega_\mu\omega^\mu + \ff4 \Omega_{\mu\nu}\Omega^{\mu\nu}  \re &+ \ff2 m_\rho^2\ivec \rho_\mu\cdot\ivec\rho^\mu -\ff4\ivec R_{\mu\nu}\cdot\ivec R^{\mu\nu} + \ff2\partial_\mu\ivec\pi\cdot\partial^\mu\ivec\pi -\ff4 F_{\mu\nu}F^{\mu\nu},\re
\scr L_I = & -\bar\psi\left[g_\sigma\sigma + g_\omega\gamma^\mu\omega_\mu + g_\rho \gamma^\mu\ivec\tau\cdot\ivec\rho_\mu - \frac{f_\rho}{2M}\sigma^{\mu\nu} \ivec\tau\cdot\partial_\nu\ivec\rho_\mu\right.\re &\hspace{2em}\left. + \frac{f_\pi}{m_\pi}\gamma_5\gamma^\mu\ivec\tau\cdot\partial_\mu\ivec\pi + e\gamma^\mu\frac{1-\tau_3}{2} A_\mu\right]\psi, \label{LI}
\end{align}
where $\Omega^{\mu\nu}= \partial^\mu\omega^\nu - \partial^\nu\omega^\mu$, $\ivec R^{\mu\nu}= \partial^\mu\ivec R^\nu - \partial^\nu\ivec R^\mu$, and $F^{\mu\nu}= \partial^\mu F^\nu - \partial^\nu F^\mu$. In the Lagrangians (\ref{Lfree}-\ref{LI}), $M$ and $m_i$ ($g_i$ or $f_i$) denote the masses (coupling constants) of (between) nucleon and mesons. In the above expressions and the following context, the arrows are used to denote isovector quantities, and the bold types for the vectors in coordinate space.

Following the standard variational procedure, the Hamiltonian $H$ can be effectively derived from the Lagrangian density $\scr L$ as,
\begin{align}\label{Hamitonian}
H&= \int d\svec x\  \bar\psi(\svec x)\lrb{-i\svec\gamma\cdot\svec\nabla + M}\psi(\svec x)\\
&+\ff2 \sum_{\phi} \int d\svec x_1 d\svec x_2\ \bar\psi(\svec x_1)\bar\psi(\svec x_2)\Gamma_\phi(1,2) D_\phi(1,2) \psi(\svec x_2)\psi(\svec x_1),\nonumber
\end{align}
where $\phi$ denotes the meson-nucleon coupling channels, namely the Lorentz scalar ($\sigma$-S), vectors ($\omega$-V, $\rho$-V and $A$-V), vector-tensor ($\rho$-VT), tensor ($\rho$-T) and pseudo-vector ($\pi$-PV) couplings.  In the Hamiltonian (\ref{Hamitonian}), the interacting vertices $\Gamma_\phi(1, 2)$ read as
\begin{subequations}\label{vertex}
\begin{align}
\Gamma_\sigs(1,2) \equiv& -g_\sigma(1)g_\sigma(2), \\
\Gamma_\omev(1,2) \equiv& +\lrs{ g_\omega \gamma_\mu}_1 \lrs{g_\omega\gamma^\mu}_2, \\
\Gamma_\rhov(1,2) \equiv& +\lrs{g_\rho\gamma_\mu\ivec\tau }_1 \cdot \lrs{ g_\rho\gamma^\mu\ivec\tau}_2, \\
\Gamma_\rhot(1,2) \equiv& +\lrs{\frac{f_\rho}{2M} \sigma_{\mu\nu}\ivec\tau \partial^\nu}_1\cdot \lrs{\frac{f_\rho}{2M} \sigma_{\mu\lambda}\ivec\tau \partial^\lambda}_2, \label{rhot}\\
\Gamma_\rhvt(1,2) \equiv& +\lrs{\frac{f_\rho}{2M}\sigma_{\mu\nu}\ivec\tau\partial^\mu}_1\cdot\lrs{g_\rho \gamma^\nu\ivec\tau}_2\re &+\lrs{g_\rho \gamma^\nu\ivec\tau}_1\cdot\lrs{\frac{f_\rho}{2M}\sigma_{\mu\nu}\ivec\tau\partial^\mu}_2, \label{rhvt}\\
\Gamma_\pipv(1,2) \equiv& -\lrs{\frac{f_\pi}{m_\pi}\ivec\tau\gamma_5\gamma_\mu\partial^\mu }_1 \cdot \lrs{\frac{f_\pi}{m_\pi}\ivec\tau\gamma_5\gamma_\nu\partial^\nu }_2, \label{pi-sv}\\
\Gamma_\couv \equiv& +  \lrs{e\gamma_\mu\frac{1-\tau_3}2}_1 \lrs{e\gamma^\mu\frac{1-\tau_3}2}_2,
\end{align}
\end{subequations}
and $D_\phi(1,2)$ are the propagators of meson and photon fields with the following Yukawa form:
\begin{align}
D_\phi=&\ff{4\pi} \frac{e^{-m_\phi\lrl{\svec x_1 -\svec x_2}}}{\lrl{\svec x_1 -\svec x_2}}, & D_{\couv} = & \ff{4\pi} \frac{1}{\lrl{\svec x_1 -\svec x_2}}.
\end{align}
It should be noticed that in deriving the Hamiltonian (\ref{Hamitonian}) we have introduced the simplifying assumption of neglecting the time component of the four-momentum carried by the mesons, which means that the meson fields are time independent. This assumption has no consequence, in the static case, on the direct (Hartree) terms while it amounts to neglecting the retardation effects for the exchange (Fock) terms \cite{Bouyssy1987}.

To provide an accurate quantitative description of nuclear systems, one also has to treat the nuclear in-medium effects of the nucleon-nucleon interactions properly, either by introducing the non-linear self-couplings of the meson fields \cite{Boguta1977, Bernardos1993, Marcos2004} or the density dependence of meson-nucleon couplings \cite{Lenske1995, Long640}. In the current framework, i.e., the density-dependent relativistic Hartree-Fock (DDRHF) theory \cite{Long640, Long2007}, the meson-nucleon coupling constants are assumed to be a function of baryon density $\rho_b$. For the isoscalar $\sigma$- and $\omega$-mesons, the density dependences of the coupling constants $g_i$ ($i=\sigma$, $\omega$) are chosen as,
\begin{align}
  g_i(\rho_b)=& g_i(\rho_0) f_i(\xi), &   f_i(\xi)= & a_i \frac{1+b_i\lrb{\xi + d_i}^2}{1+c_i\lrb{\xi + d_i}^2},
\end{align}
where $\xi=\rho_b/\rho_0$, and $\rho_0$ denotes the saturation density of nuclear matter. In addition, five constraint conditions $f_i(1)=1$, $f''_\sigma(1)=f''_\omega(1)$, and $f''_i(0)=0$ are introduced to reduce the number of free parameters. For the ones in the isovector channels, i.e., $g_\rho$, $f_\rho$, and $f_\pi$, an exponential density dependence is utilized,
\begin{align}
  g_\rho=& g_\rho(0) e^{-a_\rho\xi},&  f_\rho=& f_\rho(0) e^{-a_T\xi},&
  f_\pi =& f_\pi(0) e^{-a_\pi \xi}.
\end{align}

At the mean field level, the contributions from the Dirac sea are neglected, i.e., the widely used no-sea approximation. Consequently the HF ground state can be determined as,
\begin{align}
  |\Phi_0\rangle =& \prod_\alpha c_\alpha^\dag |0\rangle,
\end{align}
where $c_\alpha^\dag$ is the creative operator of the particle, $|0\rangle$ is the vacuum state, and the index $\alpha$ only runs over the positive energy states. With respect to the ground state $\lrlc{\Phi_0}$, the RHF energy functional can be obtained from the expectation of the Hamiltonian $H$ as,
\begin{align}
E=\lrc{\Phi_0\lrl{H}\Phi_0} \equiv \lrc{\Phi_0\lrl{T}\Phi_0} + \ff2 \sum_\phi \lrc{\Phi_0\lrl{V_\phi}\Phi_0},
\end{align}
where $T$ and $V_\phi$ denote the kinetic and potential energy parts, respectively, and the later contains two types of contributions: the direct (Hartree) $\lrc{V}_D$ and exchange (Fock) terms $\lrc{V}_E$ \cite{Bouyssy1987}.

\subsection{Relativistic representation of nuclear tensor force components}
In Ref. \cite{Jiang2014}, the relativistic formalism to identify the nuclear tensor force components hidden in the Fock terms of the meson-nucleon couplings are proposed respectively for $\pi$-PV, $\sigma$-scalar (S), $\omega$-vector (V) and $\rho$-tensor (T) couplings, and they read as,
\begin{align}
\scr H_\pipv^T = & -\ff2\lrs{ \frac{f_\pi}{m_\pi} \bar\psi  \gamma_0 \Sigma_\mu\ivec\tau \psi }_1\cdot\lrs{\frac{f_\pi}{m_\pi} \bar\psi\gamma_0 \Sigma_\nu \ivec\tau \psi}_2 D_\pipv^{T,\ \mu\nu}(1,2),\label{Wigner-tensor-pi-R}\\
\scr H_\sigs^T = & -\ff4\lrs{\frac{g_\sigma}{m_\sigma}\bar\psi \gamma_0\Sigma_\mu\psi }_1\lrs{\frac{g_\sigma}{m_\sigma}\bar\psi \gamma_0\Sigma_\nu \psi }_2 D_\sigs^{T,\ \mu\nu}(1,2),\label{Wigner-tensor-sig-R}\\
\scr H_\omev^T = & +\ff4\lrs{\frac{g_\omega}{m_\omega}\bar\psi \gamma_\lambda\gamma_0\Sigma_\mu\psi }_1 \lrs{\frac{g_\omega}{m_\omega}\bar\psi \gamma_\delta \gamma_0\Sigma_\nu \psi }_2 D_\omev^{T,\ \mu\nu\lambda\delta}(1,2),\label{Wigner-tensor-ome-R}\\
\scr H_\rhot^T = & +\ff2 \lrs{\frac{f_\rho}{2M}\psi \sigma_{\lambda\mu}\ivec\tau\psi }_1\cdot \lrs{\frac{f_\rho}{2M}\psi \sigma_{\delta\nu}\ivec\tau\psi }_2 D_\rhot^{T,\ \mu\nu\lambda\delta} (1,2),\label{Wigner-tensor-rhot-R}
\end{align}
where $\Sigma^\mu = \lrb{\gamma^5, \svec\Sigma}$, and the propagator terms $D^T$ read as,
\begin{align}
D_\phi^{T,\ \mu\nu}(1,2) = &\lrs{\partial^\mu(1)\partial^\nu(2) - \ff3 g^{\mu\nu} m_\phi^2} D_\phi (1,2)\re
&\hspace{4.8em} + \ff3 g^{\mu\nu} \delta(x_1-x_2), \label{propagator-S}\\
D_{\phi'}^{T,\ \mu\nu\lambda\delta}(1,2) = & \partial^\mu(1)\partial^\nu(2) g^{\lambda\delta}  D_{\phi'}(1,2) \re
&- \ff3 \lrb{g^{\mu\nu} g^{\lambda\delta} - \ff3 g^{\mu\lambda} g^{\nu\delta} } m_{\phi'}^2 D_{\phi'}(1,2)\re
&+ \ff3 \lrb{g^{\mu\nu} g^{\lambda\delta} - \ff3 g^{\mu\lambda} g^{\nu\delta} } \delta(x_1-x_2).\label{propagator-V}
\end{align}
In above expressions (\ref{propagator-S}-\ref{propagator-V}), $\phi$ stands for the $\sigma$-S and $\pi$-PV couplings, and $\phi'$ represents the $\omega$-V and $\rho$-T channels. For the $\rho$-V coupling, corresponding formalism $\scr H_\rhov^T$ can be obtained simply by replacing $m_\omega$ ($g_\omega$) in Eqs. (\ref{Wigner-tensor-ome-R}) and (\ref{propagator-V}) by $m_\rho$ ($g_\rho$) and inserting the isospin operator $\ivec\tau$ in the interacting index. In consistence with the theory itself, the $\mu,\nu=0$ components of the propagator terms will be omitted in practice, which amounts to neglecting the retardation effects. Transferring to the momentum space, the interaction index together with the propagator term in $\scr H_\phi^T$ ($\phi=\sigs$ and $\pipv$) can be expressed as,
\begin{align}\label{VphiT}
  V_\phi^T(\svec q) = & \ff3\frac{3\big(\gamma_0\svec\Sigma_1\cdot\svec q\big)\big(\gamma_0\svec\Sigma_2\cdot\svec q\big) -  \big(\gamma_0\svec\Sigma_1\big)\cdot\big(\gamma_0\svec\Sigma_2\big)\svec q^2}{m_\phi^2+\svec q^2},
\end{align}
and the numerator term in the right-hand side is exactly a rank-2 irreducible tensor operator similar as $S_{12}$ [see Eq. (\ref{Tensor_Wigner})]. For $\phi'=\omev$, $\rhot$ and $\rhov$, one may obtain the irreducible tensor operators with higher ranks. The $\svec q^2$ term in the numerator of Eq. (\ref{VphiT}), together with the denominator $m_\phi^2 + \svec q^2$, contributes two types of the interactions,
\begin{align}
  \frac{\svec q^2}{m_\phi^2 +\svec q^2} = & 1 - \frac{m_\phi^2}{m_\phi^2 + \svec q^2},
\end{align}
which are respectively the delta and $m_\phi^2$ terms in the propagator term (\ref{propagator-S}) if transferring back to the coordinate space.

For the uniform nuclear matter, relevant contributions to the energy density functional (EDF) from the nuclear tensor force components, namely the expectations of the proposed Hamiltonians (\ref{Wigner-tensor-pi-R}-\ref{Wigner-tensor-rhot-R}), can be derived as,
\begin{align}
E_\sigma^T =& +\ff2 \ff{(2\pi)^4} \frac{g_\sigma^2}{m_\sigma^2} \sum_{\tau_1,\tau_2} \delta_{\tau_1, \tau_2} \int p_1dp_1 p_2dp_2 \re
&\times\hat P_1\hat P_2 \lrs{\lrb{p_1^2 + p_2^2-\ff3m_\sigma^2}\Phi_\sigma - p_1p_2\Theta_\sigma }, \label{eq:tsig}\\
E_\omega^T =& +\ff{(2\pi)^4} \frac{g_\omega^2}{m_\omega^2} \sum_{\tau_1, \tau_2} \delta_{\tau_1, \tau_2} \int p_1dp_1 p_2dp_2 \re
&\times \Big\{\lrs{ \lrb{p_1^2 + p_2^2 + \ff6 m_\omega^2}\Phi_\omega - p_1p_2\Theta_\omega } \hat P_1 \hat P_2 \re
& \hspace{2 em} +\lrb{\ff4 m_\omega^2\Theta_\omega - p_1p_2} \lrb{\hat M_1\hat M_2 -1} \Big\}, \label{eq:tome}\\
E_\pi^T =& + \ff{(2\pi)^4} \frac{f_\pi^2}{m_\pi^2} \sum_{\tau_1,\tau_2}\lrb{2- \delta_{\tau_1, \tau_2} } \int p_1dp_1 p_2dp_2 \re
&\times \hat P_1\hat P_2\lrs{\lrb{p_1^2 + p_2^2-\ff3m_\pi^2}\Phi_\pi - p_1p_2\Theta_\pi  }, \label{eq:tpio}\\
E_{\rhot}^T =& +\ff2 \ff{(2\pi)^4} \frac{f_\rho^2}{M^2} \sum_{\tau_1,\tau_2}\lrb{2- \delta_{\tau_1, \tau_2} } \int p_1dp_1 p_2dp_2 \re
&\times \hat P_1\hat P_2\lrs{\lrb{p_1^2 + p_2^2-\ff3m_\pi^2}\Phi_\pi - p_1p_2\Theta_\pi}, \label{eq:trtn}
\end{align}
where $\tau_1$ and $\tau_2$ denote the third components of the isospin of nucleons. For the quantities $\Theta$, $\Phi$, and the hatted ones $\hat P$ and $\hat M$, they read as,
\begin{align}
  \Theta_\phi(p_1, p_2) = & \ln \frac{m_\phi^2 + \big (p_1 + p_2\big)^2}{m_\phi^2 + \big (p_1 - p_2\big)^2}, \\
  \Phi_\phi(p_1, p_2) = & \frac{p_1^2 + p_2^2 + m_\phi^2}{4p_1p_2} \Theta_\phi(p_1, p_2) - 1,\\
  \hat{\svec P} = & \frac{\svec p^*}{E^*}, \hspace{2em}  \hat M =  \frac{ M^*}{E^*},
\end{align}
with $\svec p^* = \svec p + \hat{\svec p} V_V$, $M^* = M+V_S$ and $E^* = E - V_0$ \cite{Bouyssy1987}, where $V_S$ is the scalar self-energy, and $V_0$ and $V_V$ are the time and space components of the vector one, respectively.
For $\rho$-V coupling, its expression can be obtained by replacing $m_\omega$($g_\omega$) and isospin factor $\delta_{\tau_1,\tau_2}$ in Eq. (\ref{eq:tome}) with $m_\rho$($g_\rho$) and $\lrb{2-\delta_{\tau_1, \tau_2}}$, respectively. With the EDFs (\ref{eq:tsig}-\ref{eq:trtn}) of the nuclear tensor forces, the corresponding contributions to the self-energies can also be obtained. Notice that the extraction of the tensor force contributions does not introduce any additional free parameters, which is exactly the advantage of the method to treat the tensor effects self-consistently.

\section{Results and Discussion}\label{Sec:Result}

Since a nuclear tensor force emerges simultaneously with the presence of Fock diagrams of meson-nucleon couplings, it is worthwhile to study its effects with the proposed relativistic representation [see Eqs. (\ref{Wigner-tensor-pi-R}-\ref{Wigner-tensor-rhot-R})]. In this study, we focus on the role played by the naturally involved tensor force components in the saturation mechanism, the EoS and the symmetry energy of nuclear matter, and the bulk properties of neutron star, using the DDRHF functionals PKA1 \cite{Long2007}, PKO1 \cite{Long640}, PKO2 and PKO3 \cite{Long2008}. Among these functionals, PKA1 has the complete RHF scheme of meson-nucleon couplings as list in Eq. (\ref{vertex}), whereas in PKO series the $\rhot$ and $\rhvt$ couplings are missing, and the $\pipv$ one is not included in PKO2, either. In order to reveal the self-consistent tensor effects in describing the nuclear matter and neutron star properties, we performed the comparison between two self-consistent calculations: one with the full EDF and the other with an EDF that drops the tensor force components. With these two self-consistent procedures, the tensor force contributions to the EDF can be completely included or excluded, respectively.

\subsection{symmetric nuclear matter}\label{Sec:Result-EOS}

\begin{table}[htbp]
\centering
\caption{ Bulk properties of symmetric nuclear matter at saturation point, i.e., the saturation density $\rho_0$ in unit of fm$^{-3}$, binding energy per particle $E/A$ and incompressibility $K$ in unit of MeV. The results are calculated by using the DDRHF functionals PKA1, PKO1, PKO2 and PKO3. The results which drop the tensor contributions (W/O) are given in the brackets for comparison.}
\label{tab:sat-bulk}
\begin{tabularx}{8.5cm}{@{\extracolsep{\fill}}cccc}
\hline\hline
            & $\rho_0$(W/O) &     $E/A$(W/O)        &  $K$(W/O) \\
\hline
    PKA1    & 0.160~(0.148) & $-$15.83~($-$14.18)   & 229.96~(203.56) \\
    PKO1    & 0.152~(0.140) & $-$16.00~($-$14.21)   & 250.24~(221.96) \\
    PKO2    & 0.151~(0.139) & $-$16.03~($-$14.31)   & 249.60~(222.65) \\
    PKO3    & 0.153~(0.140) & $-$16.04~($-$14.22)   & 262.47~(229.82) \\
\hline\hline
\end{tabularx}
\end{table}

Table \ref{tab:sat-bulk} shows the bulk properties of symmetric nuclear matter at saturation point, namely the saturation density $\rho_0$ (fm$^{-3}$), the binding energy per nucleon $E/A$ (MeV) and the incompressibility $K$ (MeV). To reveal the tensor effects in determining the saturation mechanism, the values in the brackets are the results extracted from the calculations which drop the tensor contributions. With the full DDRHF functionals which have the nuclear tensor force components involved in the Fock diagrams automatically, the saturation points have been well established as the saturation density $\rho_0$ $\thicksim 0.16$ fm$^{-3}$ and the binding energy $E/A$ $\thicksim-16$ MeV, and both are in a good agreement with the empirical values. As generally expected, the nuclear tensor force presents tiny contributions to the energy functional indeed. While the saturation mechanism is disturbed essentially, if removing the tensor force contributions from the DDRHF functionals. As shown in Table \ref{tab:sat-bulk}, the saturation densities become 0.012 fm$^{-3}$ smaller due to the dropping of the nuclear tensor force components and such reduction almost accounts for 8\% of the saturation density. Consistently the changes of 1.7 MeV are found on the binding energy per nucleon $E/A$. For the incompressibility $K$ that has wide ranges in theoretical predictions \cite{Stone2014}, the tensor effects enhance distinctly the $K$ values by 26 $\thicksim$ 33 MeV.

\begin{figure}[floatfix]\centering
\includegraphics[width = 0.45\textwidth]{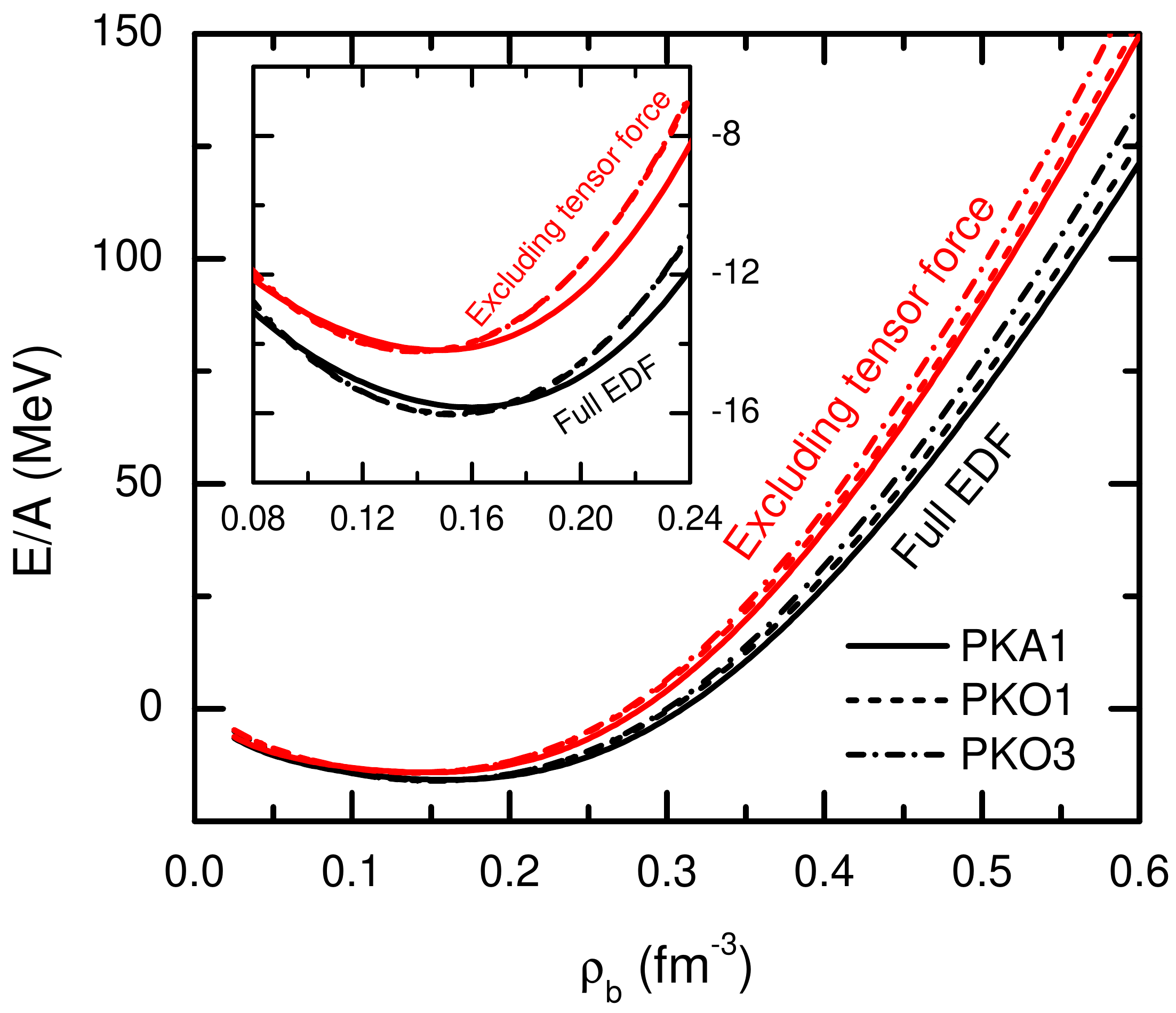}
\caption{(Color Online) The binding energy per nucleon $E/A$ (MeV) of symmetric nuclear matter as a function of the baryonic density $\rho_b$. The results are calculated with the full EDFs determined by the DDRHF functionals PKA1, PKO1 and PKO3, as compared to those dropping the tensor force contributions (red lines). The subset shows the results around $\rho_0$ with smaller scale.}
\label{fig:EoS}
\end{figure}

It should be noticed that the candidates of nuclear tensor-related observables, such as the nuclear spin-isospin resonances \cite{Bai2011PRC} and single-particle shell evolution \cite{Colo2007}, were not utilized in parameterizing the DDRHF functionals PKA1 \cite{Long2007} and PKO series \cite{Long640}. Even though, as seen from Table \ref{tab:sat-bulk}, the nuclear saturation mechanism is influenced fairly distinctly by the natural tensor force components in the DDRHF functionals. From Eqs. (\ref{eq:tsig}--\ref{eq:trtn}), these tensor contributions to the energy functional depend on the momentum $p$, and gradual enhancements on the tensor EDFs are therefore predictable at high-density region, as well demonstrated in Fig. \ref{fig:EoS}. The black lines in the Fig. \ref{fig:EoS} are the EoSs of symmetric nuclear matter calculated with the full DDRHF functionals PKA1, PKO1 and PKO3, and the red lines correspond to the relevant calculations which drop the tensor force components. Comparing the calculations with the full DDRHF functional and those dropping the tensor terms, it seems that in the low-density region ($\rho_b \lesssim \rho_0$), the nuclear tensor force does not change much the EoSs. If concentrating on the density region $\rho_b\sim\rho_0$ with smaller scale, the deviations between two types of the calculations are still remarkable as seen from the subset in Fig. \ref{fig:EoS}. Qualitatively it can be easily justified that the presence of the tensor terms in the full DDRHF functionals increase the curvatures of the EoSs at $\rho_b=\rho_0$, i.e., the values of the incompressibility $K$ are enhanced by the tensor effects. In the supranuclear density region, the nuclear tensor force presents much more distinct effects, which contributes about 40 MeV to the energy functional and makes the EoSs softer.

\subsection{symmetry energy}\label{Sec:Result-Esym}

The symmetry energy and its density-dependent behavior play a crucial role in understanding the properties of neutron-rich nuclei, isospin asymmetric nuclear matter, and neutron stars. Despite much efforts were devoted by the experimental and theoretical researchers, the density behavior of symmetry energy at supranuclear density region is still not well constrained. Theoretically very different high-density behaviors of symmetry energy were predicted by various models, varying from extreme soft to very stiff ones \cite{Brown2000, Feng2012, Tsang2012}. Recently, some studies were performed to reveal the tensor effects on the density dependence of the symmetry energy \cite{Baoan2010, Vidana2011}.

\begin{table}[htbp]
\centering
\caption{The symmetry energy $J$ together with its slope $L$ and curvature $K_{\rm sym}$ obtained by the calculations of DDRHF with PKA1, PKO1, PKO2 and PKO3. The results which drop the tensor contributions (W/O) are listed in the brackets for comparison.  All values are in unit of MeV.}
\label{tab:sat-Esym}
\begin{tabularx}{8.5cm}{@{\extracolsep{\fill}}cccc}
\hline\hline
            & $J$(W/O)       &     $L$(W/O)      & $K_\text{sym}$(W/O) \\
\hline
    PKA1    & 36.02~(35.95)  & 103.50~ (115.49)  & 212.90~(317.31) \\
    PKO1    & 34.37~(33.50)  & ~97.70~ (101.66)  & 105.85~(158.87)  \\
    PKO2    & 32.49~(31.73)  & ~75.93~ ~(81.12)  & ~77.51~(128.77)  \\
    PKO3    & 32.98~(32.26)  & ~83.00~ ~(88.91)  & 116.43~(176.39)  \\
\hline\hline
\end{tabularx}
\end{table}

\begin{figure}[htbp]\centering
\includegraphics[width = 0.45\textwidth]{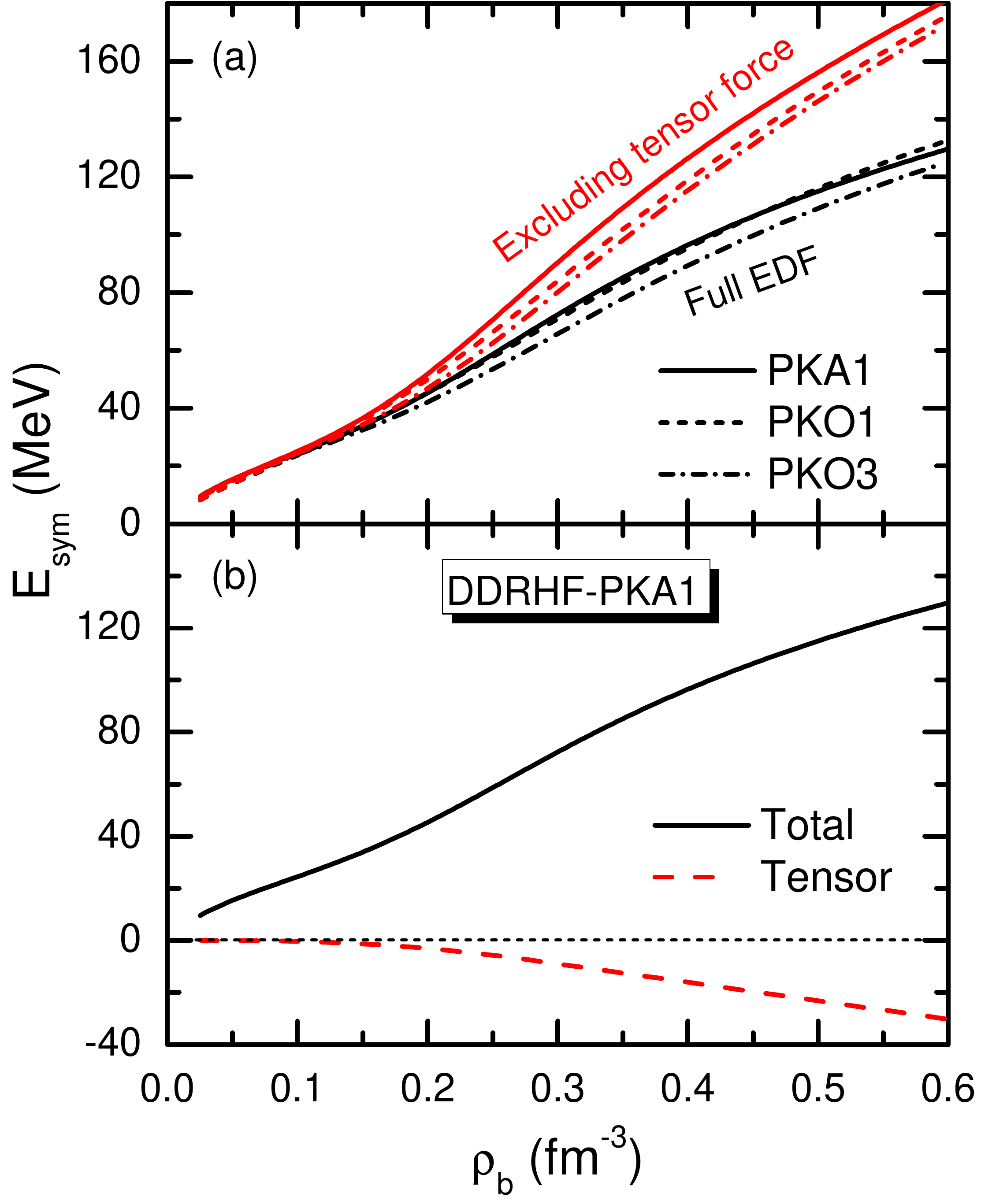}
\caption{(Color Online) The symmetry energy (MeV) of nuclear matter as a function of baryon density $\rho_b$ (fm$^{-3}$).  The results are calculated with the full EDFs determined by the DDRHF functionals PKA1, PKO1 and PKO3, as compared to those dropping the tensor force contributions (red lines). The lower plot shows the symmetry energy (solid line) and the contributions only from the tensor terms (dashed line) taking PKA1 as an example.}
\label{fig:Esym}
\end{figure}

\begin{figure}[floatfix]\centering
\includegraphics[width = 0.45\textwidth]{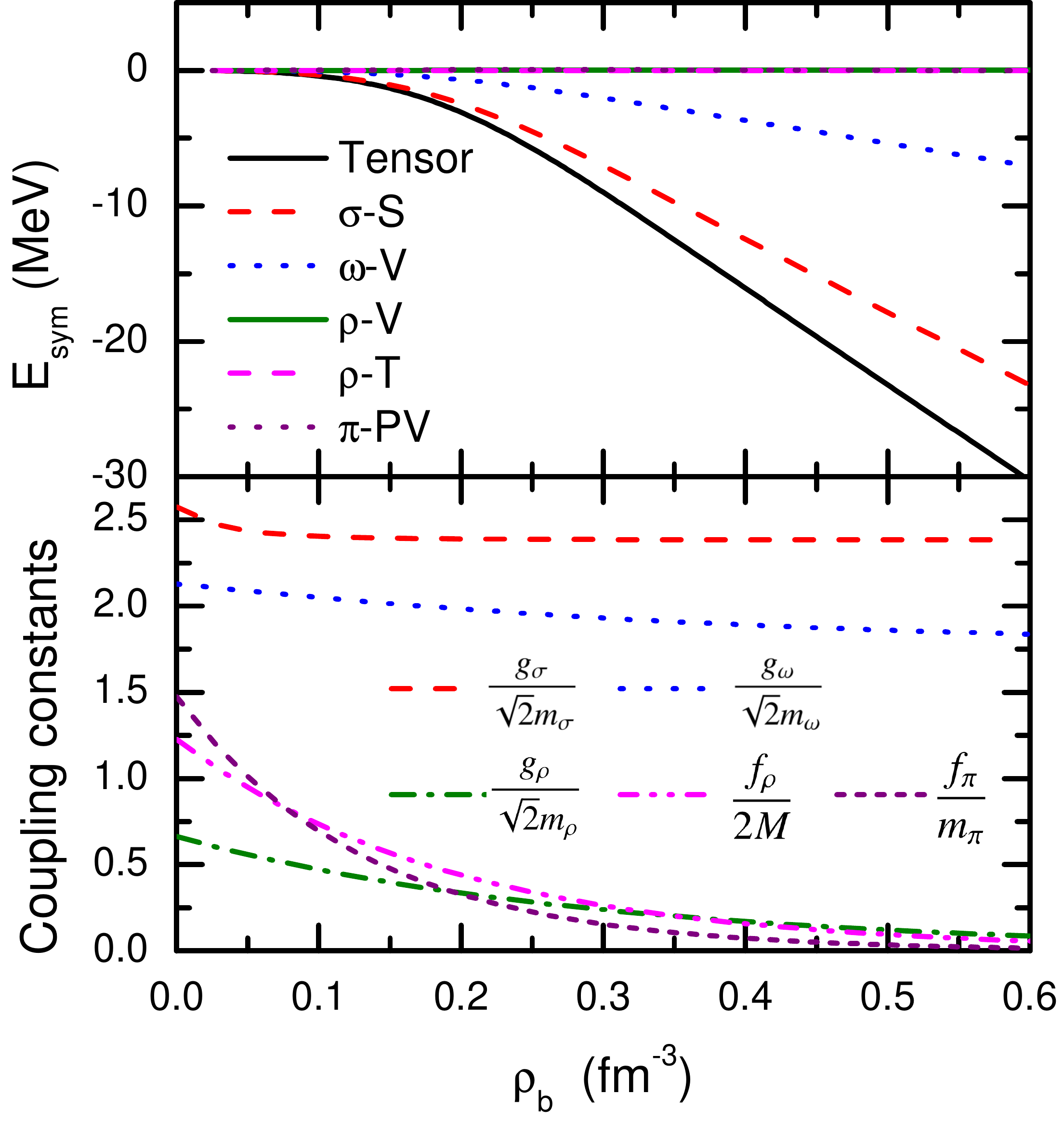}
\caption{(Color Online)  Contributions to the symmetry energy from the tensor force components in various meson-nucleon coupling channels [plot (a)] and the tensor coupling constants [plot (b)] as functions of baryon density $\rho_b$ (fm$^{-3}$). The results are extracted from the calculations of DDRHF with PKA1.}
\label{fig:O3-Esym-cctm}
\end{figure}

Table \ref{tab:sat-Esym} shows the symmetry energy $J$ with its slope $L$ and curvature $K_{\rm sym}$, and those extracted from the calculations dropping the tensor contributions are given in the brackets for comparison. For the symmetry energy $J$ at saturation density, the subtractions of the tensor contributions bring very tiny changes whereas both the slope $L$ and curvature $K_{\rm sym}$ increase fairly distinctly. Similar as the results shown in Table \ref{tab:sat-bulk}, the tensor effects on the symmetry energy $J$ with its slope $L$ and curvature $K_{\rm sym}$ are not so notable. Such results are closely connected with the nature of relativistic EDFs of the tensor force components (\ref{eq:tsig} -- \ref{eq:trtn}) which essentially depend on the momentum carried by the nucleons. In the low-density region associated with the low momentum, the nuclear tensor force shows little impact on the nuclear matter properties, and with the density increasing that is equivalent to increasing the momentum $p$ the tensor effects may become remarkable.

In Fig. \ref{fig:Esym} the symmetry energies calculated with the DDRHF functionals PKA1, PKO1 and PKO3 are shown as a function of baryon density $\rho_b$. To reveal the tensor effects, Figure \ref{fig:Esym}(a) presents the comparison between the calculations with the full functionals and those dropping the tensor terms, and using the DDRHF functional PKA1, Figure \ref{fig:Esym}(b) shows the tensor contributions to the symmetry energy. Consistent with the results in Table \ref{tab:sat-Esym}, it is found from Fig. \ref{fig:Esym} that the withdraw of the  tensor force contributions does not bring distinct changes on the symmetry energy at subsaturation density region. However, with the density increasing, the tensor effects on the symmetry energy become notable due to the fact that the relativistic EDFs [see Eqs. (\ref{eq:tsig}-\ref{eq:trtn})] of the tensor forces depend on the momentum $p$ essentially. Compared to the calculations which drop the tensor terms [red lines in Fig. \ref{fig:Esym} (a)], the symmetry energies at supranuclear density region are fairly softened by the tensor effects. This is well demonstrated by the tensor contributions to the symmetry energy in Fig. \ref{fig:Esym} (b) which are negative and counteract about 20\% of the contributions from the other channels at high density. It is worthwhile to mention that the tensor force components are naturally introduced with the presence of the Fock diagrams in the DDRHF functionals. Hence the current results provide a self-consistent explanation for the tensor effects on the density dependence of the symmetry energy.

Furthermore with the relativistic EDFs (\ref{eq:tsig} -- \ref{eq:trtn}), the tensor contributions to the symmetry energy from different channels can be extracted, namely the $\sigma$-S, $\omega$-V, $\rho$-V, $\rho$-T and $\pi$-PV couplings as shown in Fig. \ref{fig:O3-Esym-cctm} (a). Using the DDRHF functional PKA1, it is clear that the tensor component in the $\sigs$ coupling channel dominates the tensor contributions to the symmetry energy, followed by the $\omev$-couplings, while those in $\pi$- and $\rho$- exchanges are close to zero. This result can be well understood from the tensor coupling constants shown in Fig. \ref{fig:O3-Esym-cctm}(b), namely $g_\sigma/\lrb{\sqrt 2 m_\sigma}$, $g_\omega/\lrb{\sqrt2m_\omega}$, $g_\rho/\lrb{\sqrt{2} m_\rho}$, $f_\rho/\lrb{2M}$ and $f_\pi/m_\pi$ in the relativistic formalism (\ref{Wigner-tensor-pi-R}-\ref{Wigner-tensor-rhot-R}). It is seen that the tensor coupling constants in $\sigs$ and $\omev$ channels tend to certain values at high density, whereas due to the exponential density-dependent behavior of $g_\rho$, $f_\rho$ and $f_\pi$, those from the isovector $\rho$-V, $\rho$-T and $\pi$-PV channels vanish at the supranuclear density region where the tensor effects become notable.

\subsection{neutron star}\label{Sec:Result-NS}

In understanding the cooling mechanism of neutron stars, the proton fraction $x = \rho_p/\lrb{\rho_n+\rho_p}$ is a key quantity which carries significant information of the EoS of asymmetric nuclear matter. By emitting thermal neutrinos through the direct Urca (DU) processes $n \to p+e^-+\bar \nu_e$ and $p+e^-\to n+\nu_e$, the stars would cool rapidly. If the proton fraction goes beyond a threshold value $x^\text{DU}$, the DU process works. Following the triangle inequality for momentum conservation and charge neutrality condition \cite{Lattimer1991, Klahn2006}, it is easy to obtain the threshold of the proton fraction $x^\text{DU}$ as $11.1\%\leq x^\text{DU} \leq 14.8\%$.

\begin{figure}[floatfix]\centering
\includegraphics[width = 0.45\textwidth]{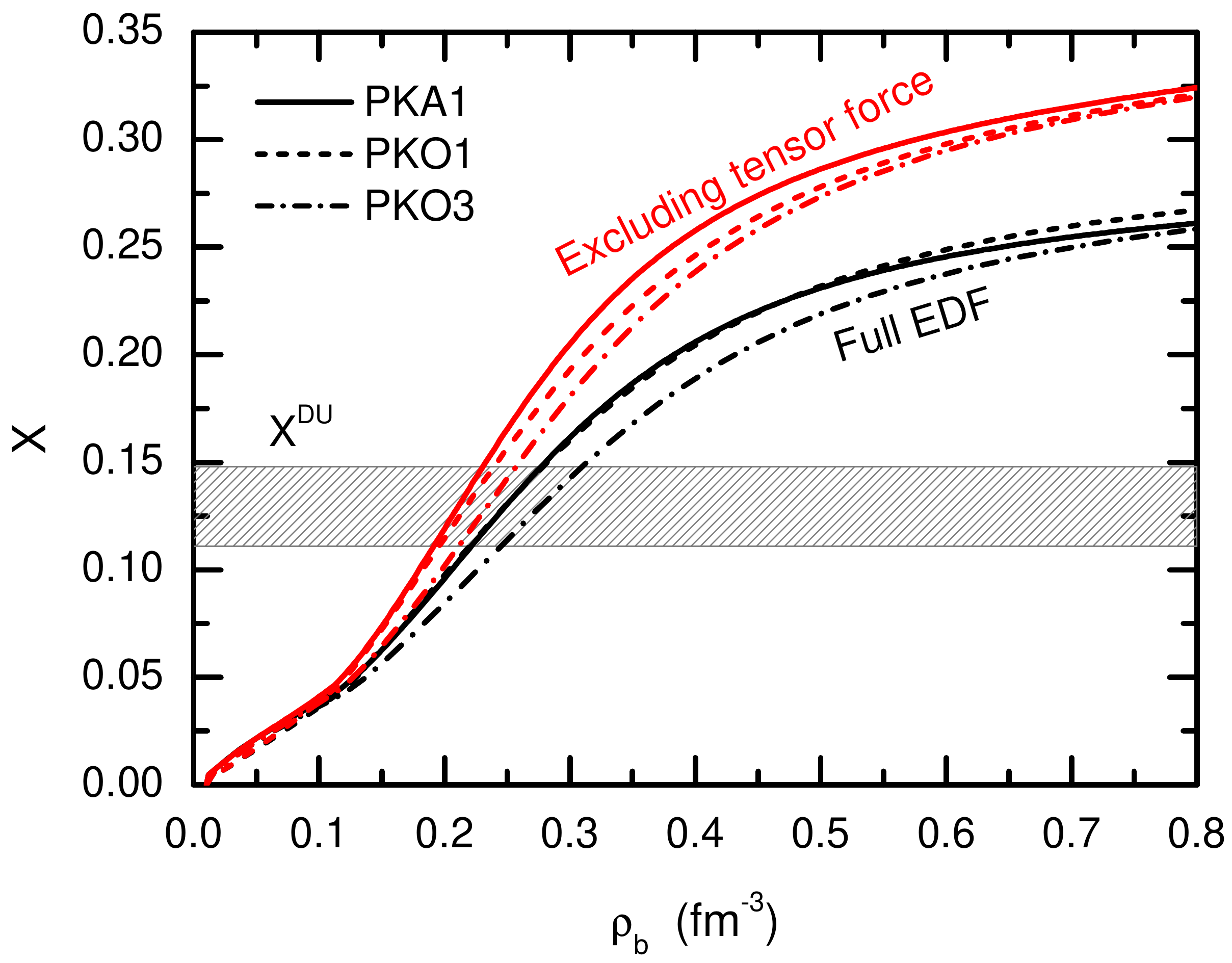}
\caption{(Color Online) Proton fraction $x=\rho_p/(\rho_p+\rho_n)$ of neutron star matter as a function of baryon density $\rho_b$ (fm$^{-3}$). The results are calculated with the full EDFs determined by the DDRHF functionals PKA1, PKO1 and PKO3, as compared to those dropping the tensor force components (red lines). The shadow area corresponds to the threshold values $11.1\%\leqslant x^\text{DU}\leqslant 14.8\%$ for the occurrence of the direct Urca process.}
\label{fig:O3-FractP}
\end{figure}

Within the density range of static and $\beta$-equilibrium neutron star matter, the proton fractions $x$ are shown as functions of baryon density $\rho_b$ in Fig. \ref{fig:O3-FractP} and the results are extracted from the calculations with the DDRHF functionals PKA1, PKO1 and PKO3, as compared to those dropping the tensor force components. It is seen from Fig. \ref{fig:O3-FractP} that the density-dependent behaviors of the proton fraction $x$ are also softened with the presence of the nuclear tensor force components in the DDRHF functionals, consistent with the systematics of the symmetry energy in Fig. \ref{fig:Esym}(a). For a given $x^{DU}$, it corresponds to a threshold density $\rho^\text{DU}$ of the DU process occurring that relies on the symmetry energy. Once when the central density $\rho_c$ of a neutron star exceeds the threshold density $\rho^\text{DU}$, the star will cool rapidly via the DU processes. One can see that the threshold density $\rho^\text{DU}$, determined by the DDRHF calculations with the full EDFs which contain the tensor force components in the Fock diagrams, are higher than those dropping the tensor terms. Such result indicates that the nuclear tensor force is unfavourable for the occurrence of the DU process. Considering the well-known fact that the occurrence of the DU process is not supported by the modern observational soft X-ray data in the cooling curve, it seems that the predictions with nuclear tensor force are in better agreement with the observations.

\begin{figure}[floatfix]\centering
\includegraphics[width = 0.45\textwidth]{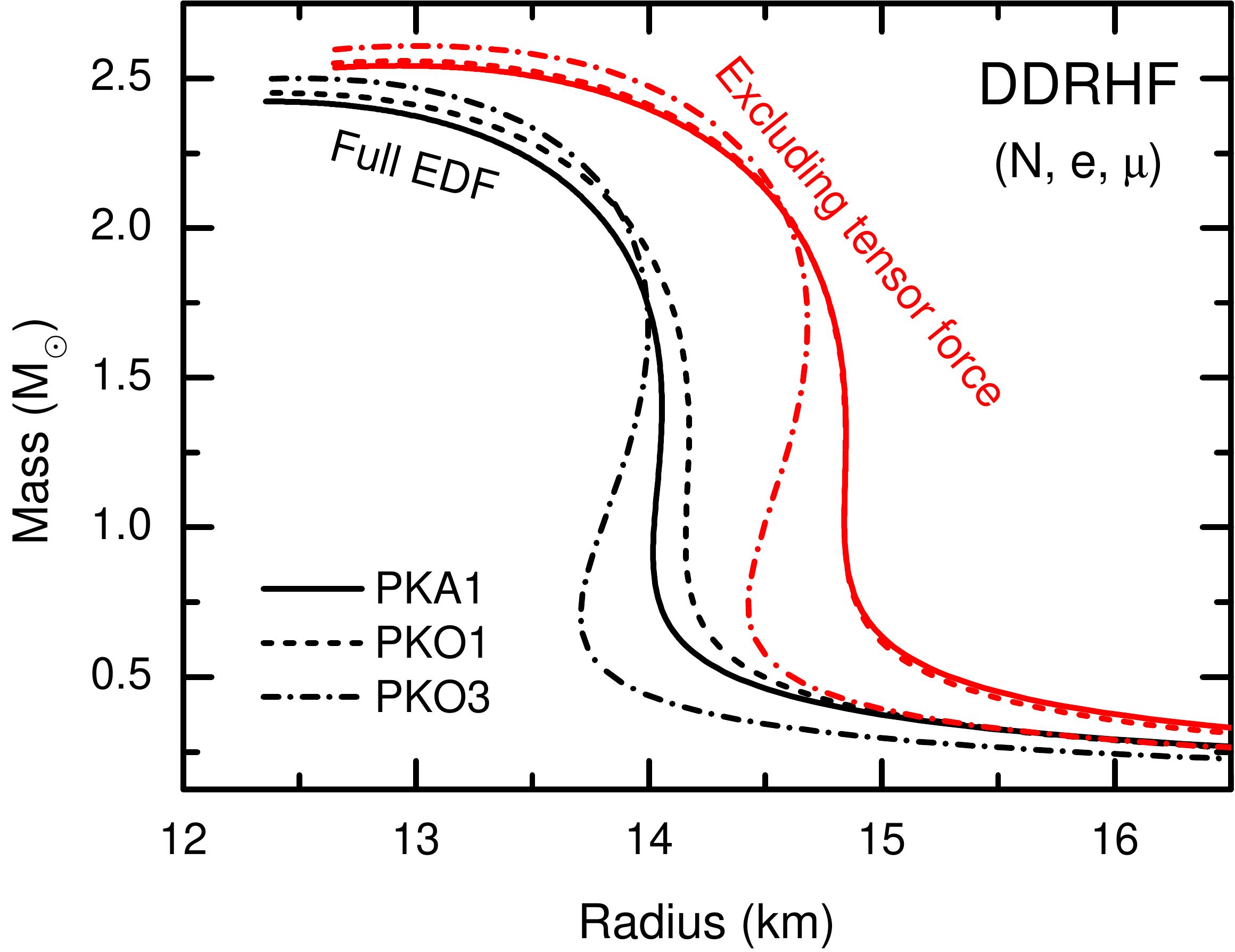}
\caption{(Color Online) Mass-radius relations of the neutron stars. The results are calculated by using the DDRHF functionals PKA1, PKO1 and PKO3, as compared to those dropping the tensor terms (red lines). }
\label{fig:TOV}
\end{figure}

\begin{table}[htbp]
\centering
\caption{The radius (km) and central density $\rho_c$ (fm$^{-3}$) of  the neutron stars with 1.4$M_\odot$ (upper panel) and the ones with $M_{\rm max}$ (lower panel). The results are calculated by using the DDRHF functionals PKA1, PKO1, PKO2 and PKO3 (W/T), as compared to those dropping the tensor terms (W/O).}
\label{tab:MR}
\begin{tabularx}{7.5cm}{@{\extracolsep{\fill}}c|rr|rr|rr}
 \hline\hline
   \multicolumn{1}{l|}{\multirow{2}{*}{$1.4M_\odot$}} & \multicolumn{2}{c|}{$M(M_\odot)$}    & \multicolumn{2}{c|}{$R(\text{km})$}   & \multicolumn{2}{c}{$\rho_c(\text{fm}^{-3})$} \\
   \cline{2-7}
   & \multicolumn{1}{c}{W/T}    & \multicolumn{1}{c|}{W/O}  & \multicolumn{1}{c}{W/T}   & \multicolumn{1}{c|}{W/O}     & \multicolumn{1}{c}{W/T}  & \multicolumn{1}{c}{W/O} \\
    \hline
    PKA1  & 1.40 & 1.40 & 14.06 & 14.84 & 0.31 & 0.27 \\
    PKO1  & 1.40 & 1.40 & 14.17 & 14.84 & 0.31 & 0.27 \\
    PKO2  & 1.40 & 1.40 & 13.79 & 14.44 & 0.32 & 0.28 \\
    PKO3  & 1.40 & 1.40 & 13.96 & 14.64 & 0.31 & 0.27 \\
\hline
   \multicolumn{1}{l|}{\multirow{2}{*}{$M_{\rm max}$}}& \multicolumn{2}{c|}{$M(M_\odot)$}    & \multicolumn{2}{c|}{$R(\text{km})$}   & \multicolumn{2}{c}{$\rho_c(\text{fm}^{-3})$} \\
   \cline{2-7}
   & \multicolumn{1}{c}{W/T}   & \multicolumn{1}{c|}{W/O}   & \multicolumn{1}{c}{W/T}   & \multicolumn{1}{c|}{W/O}  & \multicolumn{1}{c}{W/T}   & \multicolumn{1}{c}{W/O} \\
 \hline
    PKA1 & 2.42 & 2.54 & 12.35 & 12.93 & 0.81 & 0.74 \\
    PKO1 & 2.45 & 2.56 & 12.42 & 12.94 & 0.80 & 0.74 \\
    PKO2 & 2.46 & 2.56 & 12.30 & 12.82 & 0.81 & 0.74 \\
    PKO3 & 2.50 & 2.61 & 12.49 & 13.01 & 0.78 & 0.72 \\
\hline\hline
\end{tabularx}
\end{table}

Fig. \ref{fig:TOV} shows the mass-radius relation of neutron stars calculated with the DDRHF functionals PKA1, PKO1 and PKO3, and those dropping the tensor terms (in red lines) are also shown for comparison. It is found that the curves of mass-radius relation of neutron star are collectively shifted rightward about 0.8 km with the dropping of the tensor terms. In general, larger neutron star radius corresponds to a stiffer density-dependent symmetry energy. Combined with the results in Fig. \ref{fig:Esym} (a), it can be concluded that the tensor effects on the mass-radius relation of neutron star and on the density dependence of symmetry energy are congruous with each other. Due to the fact that the tensor EDFs depend on the momentum $p$ essentially [see Eqs. (\ref{eq:tsig}-\ref{eq:trtn})], fairly distinct tensor effects are therefore observed on the mass-radius relation of neutron star.

Table \ref{tab:MR} lists the radii $R$ (km) and central densities $\rho_c$ (fm$^{-3}$) of the canonical neutron stars with  1.4$M_\odot$ (upper panel) and the ones with the maximum mass limits $M_{\rm max}$ (lower panel). The results are calculated with the DDRHF functionals (W/T) and those dropping the tensor terms (W/O). For the canonical neutron stars, the radii $R$ are reduced about $0.5$ km and the central densities $\rho_c$ become larger, as compared to the calculations dropping the tensor terms. That is, the presence of the nuclear tensor force leads a neutron star to be more compact. From the lower panel of Table \ref{tab:MR}, one can also find similar systematical changes due to the nuclear tensor force.

\section{Conclusion}\label{Sec:Sum}

With the relativistic representation of the nuclear tensor forces that originate from the Fock diagrams of the meson-nucleon coupling, we studied the self-consistent tensor effects on the saturation mechanism, the equation of state, the density-dependent behavior of the symmetry energy and the neutron star properties. Within the density-dependent relativistic Hartree-Fock (DDRHF) theory, two types of the calculations were performed to reveal the tensor effects, i.e., the ones with the full DDRHF functional and those dropping the tensor terms. It is found that if removing the tensor force components in the DDRHF functionals the saturation mechanism of nuclear matter is notably influenced. Due to the fact that the tensor EDFs depend on the momentum essentially, the tensor effects become more and more distinct with the density increasing. Due to the naturally involved tensor force components in the Fock diagrams, the density-dependent behavior of the symmetry energy is fairly softened and consequently it leads neutron stars to be more compact. Moreover, for the direct Urca (DU) process that cools the neutron star rapidly, the threshold density is raised by the nuclear tensor force. Finally we would like to emphasize that different from other nuclear functionals such as the Skyrme+Tensor methods, the nuclear tensor force is included automatically with the presence of the Fock diagrams in DDRHF, and therefore the current scheme paves a self-consistent way to explore the tensor effects on the nuclear matter system.

\section*{ACKNOWLEDGMENTS}
This work is partly supported by the National Natural Science Foundation of China under Grant Nos. 11375076 and 11405223, the Specialized Research Fund for the Doctoral Program of Higher Education under Grant No. 20130211110005, and the Youth Innovation Promotion Association of Chinese Academy of Sciences.


\end{document}